\documentclass[english, a4paper, 12pt, fleqn]{article}
\usepackage[T1]{fontenc}
\usepackage[utf8]{inputenc}
\usepackage[english]{babel}
\usepackage{amsmath}
\usepackage{amssymb}
\usepackage{graphicx}

\newcommand\be{\begin{equation}}
\newcommand\bea{\begin{eqnarray} \nonumber }
\newcommand\ee{\end{equation}}
\newcommand\eea{\end{eqnarray}}

\begin{document}
\title{\Large Agnostic Risk Parity:\\ Taming Known and Unknown-Unknowns}
\author{Raphael Benichou, Yves Lemp\'eri\`ere, Emmanuel S\'eri\'e, \\
Julien Kockelkoren, Philip Seager, \\
Jean-Philippe Bouchaud \& Marc Potters \\
{\small{Capital Fund Management}} \\
{\small{23 rue de l'Universit\'e, 75007 Paris, France}}}

\date{}

\maketitle

\begin{abstract}
Markowitz' celebrated optimal portfolio theory generally fails to deliver out-of-sample diversification. In this note, we propose a new portfolio construction strategy 
based on {\it symmetry arguments} only, leading to ``Eigenrisk Parity'' portfolios that achieve equal realized risk on all the principal components of the covariance matrix. This holds true
for any other definition of uncorrelated factors. 
We then specialize our general formula to the most agnostic case where the indicators of future returns are assumed to be uncorrelated and of equal variance. This ``Agnostic Risk Parity'' (AGP)
portfolio minimizes unknown-unknown risks generated by over-optimistic hedging of the different bets. AGP is shown to fare quite well when applied to standard technical strategies such as
trend following.
\end{abstract}

\section*{Introduction}

Diversification is the mantra of rational investment strategies. Harry Markowitz proposed a mathematical incarnation of that mantra which is common lore in the professional world. 
Unfortunately, the practical implementation of Markowitz' ideas is fraught with difficulties and yields very disappointing results. 
This has been known for long, with many papers attempting to identify its flaws and suggesting remedies \cite{Black,Michaud,Tobam,Roncalli,Meucci,Meucci2}. 
The most important problems are well understood: the optimally diversified Markowitz portfolio often ends up -- somewhat paradoxically -- 
being very concentrated on a few assets only, which inevitably leads to disastrous out-of-sample risks. 
The optimal portfolio is also unstable in time and sensitive to small changes in parameters and/or expected future gains. In the face of these difficulties, two distinct branches of research have emerged. 

The first one concerns the determination of the covariance matrix of the $N$ different assets eligible in the portfolio, for example all the stocks belonging to a given index. This covariance matrix is specified by a large number of entries ($N \times (N+1)/2$) for which only a limited amount of data is available ($N \times T$, where $T$ is the length of the time series at one's disposal). When $T$ is not extremely large compared to $N$, the empirically determined covariance matrix is highly unreliable and leads to severe instabilities when used in the Markowitz optimisation program. Recently, some powerful mathematical tools have been proposed to optimally ``clean'' the empirical covariance matrix, leading to a very significant improvement in the efficiency of Markowitz diversification using the so-called ``Rotationally Invariant Estimator'' (RIE); for a short review see \cite{RIE} and refs. therein. 

Another crucial step, of course, is to specify a list of expected returns for each asset. These expected returns result either from quantitative signals 
(such as trend following) or from other form of analysis (quantitative or subjective). These signals are usually extremely noisy and unreliable, so one should rather 
speak, as we will do below, of {\it indicators}, i.e. possibly suboptimal and biased predictions of future returns. 

Once all this is done, however, a time-worn but fundamental problem remains \cite{Keynes,Taleb}. Even when sophisticated statistical tools can adequately deal with {\it risk}, they cannot handle {\it uncertainty}, i.e. the intrinsic propensity of financial markets to behave in a way that is not consistent with prior probabilities. For example, although the future ``true'' covariance matrix is often reasonably close to the cleaned (RIE) covariance matrix, correlations can also suddenly shift to a new regime that was never observed in the past. This is in fact worse for expected returns that are even more exposed to unknown-unknowns than volatility or correlations. One therefore needs an extra layer of control, beyond Markowitz' optimisation, that acts as a safeguard against statistically unexpected events. 

This is what the second strand of research mentioned above attempts to address. The idea is to add to the standard risk-return objective function some extra penalty terms that enforce diversification, typically in the form of generalized Herfindahl indices or entropy functions \cite{Aguilar,Meucci,Wagner}. This has led to important breakthroughs, such as the concept of ``Maximally Diversified Portfolios'' (MDP) \cite{Tobam}, or more recently, of ``Principal Risk Parity Portfolios'' (PRP) (with several variations on this theme, see Refs. \cite{PP,Meucci2,MaxEnt,Lohre,Bailey}). 

\section*{Diversification and Isotropy}

Although interesting, there is a hidden assumption in these penalty terms that is far from neutral, which is the choice of the assets one considers as ``fundamental'', among which risk should be as diversified as possible in the portfolio. These assets are chosen to be physical stocks for MDP's or the principal components of the correlation matrix in the case of PRP's. In the case of long-only portfolios and traditional asset management, the choice of physical assets as the natural ``basis'' for portfolio construction might be reasonable. But for -- say -- a portfolio of futures contracts with long and short positions, any linear combination of these assets is a priori feasible (at least within some overall leverage constraint). In mathematical terms, one can ``rotate'' the natural asset basis into any a priori equivalent one. The point, however, is that a maximally {\it diversified} portfolio  in one basis can in fact become maximally {\it concentrated} in another! Take for example a portfolio of stocks with equal weights $w_i=1/N$ on all $N$ stocks. From the point of view of the (neg)entropy $S = \sum_{i} w_i \ln w_i$ or of the Herfindahl index $H = \sum_{i} w_i^2$, this is clearly optimal. But since the leading risk factor associated with the correlation matrix is itself very close to an equi-weighted allocation on all stocks, a rotation onto the principal component basis $\alpha$ leads to the worse possible values for both the entropy and the Herfindahl index. In other words, the very concept of maximal diversification is not invariant under a redefinition of the assets considered as ``fundamental''. Another vivid example of the arbitrariness in the definition of fundamental assets is provided by the interest rate curve or more generally, of contracts with different maturities. Should one consider the physical contracts, or only one of them and all associated calendar spreads?  

Are there special directions in asset space that play a special role? Can one unambiguously identify risk factors that are more fundamental than others? This is an old problem in quantitative finance, with a long list of papers attempting to identify these factors, in particular in the equity space. However, as recently reviewed by Roll \cite{Roll}, there is no consensus on this point. If risk is associated to volatility (or variance), then the problem is in fact completely degenerate or, using mathematical parlance, {\it isotropic}. 

To make this clear, let us consider asset returns $r_i$ ($i=1,\dots,N$) as random variables with zero mean\footnote{Here and below, we assume that any non zero average return (coming for example from predictive signals) is small compared to the volatility, and can be neglected in our discussion. Still, of course, this non zero average returns is what motivates the portfolio construction to start with!} and (true) covariance matrix ${\bf{C}}$, with ${\bf{C}}_{ij} = \mathbb{E}[r_i r_j]$. One can then build $N$ linear combinations of assets such that their returns $\widehat r_\alpha$ are all uncorrelated and of unit variance. But this choice is not unique: in fact, any further rotation\footnote{What we call rotations in this paper in fact includes both proper and improper rotations, i.e. rotations plus inversions.} in the space of assets (i.e. an orthogonal combination of the synthetic assets returns $\widehat r_\alpha$) leads to another set of uncorrelated, unit variance assets -- see below. Among this infinite choice of potential ``factors'', is there any one that stands out, that would justify applying a maximum diversification criterion among these special assets? This is the path followed in, e.g. \cite{Torsion}, where the further notion of ``Minimum Torsion Bets'' was introduced. 

\section*{Symmetries}

We want to propose here a related, but different route based on symmetry arguments, which fully exploits rotation and dilation invariance at the level of indicators as well as at the level of returns. First, let us 
note that one can rescale the returns of each asset $i$ by an arbitrary factor without changing the portfolio allocation problem. Investing $1$ in a stock is the same as investing $\frac12$ on a fictitious ``2-stock'' contract, with twice the returns as the original stock. So we can always choose to work with returns with unit variance, a choice that we will make henceforth. In this case, the covariance 
matrix ${\bf{C}}$ is in fact the correlation matrix between stocks. Now, the linear transformation
\be\label{hatr}
\widehat r_i = \sum_j \left({\bf{C}}^{-1/2}\right)_{ij} r_j
\ee
is such that $\mathbb{E}[\widehat r_i \widehat r_j] = \delta_{ij}$, i.e., to a set of uncorrelated assets.
Here ${\bf{C}}^{-1/2}$ is defined as the positive-definite square root of ${\bf{C}}$, namely:
\be
{\bf C}^{-1/2}=\sum_a \frac{1}{\sqrt{\lambda_a}} \,\, {\bf u}_a {\bf u}_a^T,
\ee
where $\lambda_a$ and ${\bf u}_a$ are the eigenvalues and eigenvectors of ${\bf C}$.
This is the meaning we will give throughout this paper to the square-root of a symmetric matrix. 
As noted above, there is a large degeneracy in the construction of the set of uncorrelated assets: any rotation of ${\bf \widehat r}$ would do.
A natural choice at this point is to insist that the $\hat r_i$'s are ``as close as possible'' to the original normalized returns, so that the financial intuition about the resulting synthetic assets is preserved (to wit, $\widehat{\text{SPX}} \approx {\text{SPX}}$).
This is the case for the $\hat r_i$'s defined in Eq. (\ref{hatr}) (see Appendix for a proof of this statement)\footnote{The choice of normalization for the returns ${\bf r}$ is important here. Indeed working with non-normalized returns would lead to a different result for ${\bf \widehat r}$. The choice we made is in line with our isotropy assumption.}. 
%
%
% However, as noted above, there is a large degeneracy in the construction of ${\bf{C}}^{-1/2}$ itself, which is only defined up to an arbitrary rotation $\mathbb{R}$. A natural choice at this point is to insist that the $\hat r_i$'s defined in Eq. (\ref{hatr}) are ``as close as possible'' to the original normalized returns, so that the financial intuition about the resulting synthetic assets is preserved (to wit, $\widehat{\text{SPX}} \approx {\text{SPX}}$). This constraint fixes the rotation $\mathbb{R}$ in such a way that 
% ${\bf{C}}^{-1/2}$ is given by 
% \be
% {\bf C}^{-1/2}=\sum_a \frac{1}{\sqrt{\lambda_a}} \,\, {\bf u}_a {\bf u}_a^T,
% \ee
% where $\lambda_a$ and ${\bf u}_a$ are the eigenvalues and eigenvectors of ${\bf C}$ (see Appendix for a proof of this statement). This is the meaning we will give throughout this paper to the square-root of a symmetric matrix. 

The same construction can be applied for statistical indicators of future returns that we call $p_i$, $i=1, \dots, N$. We insist that $p_i$ is not necessarily the ``true'' expectation value of the future $r_i$, but simply the best guess of the investor based on his information/skill set/biases, etc. A standard example considered below is a trend indicator based on a moving average of past returns, but any quantitative indicator based on information or intuition would do.  These indicators fluctuate in time and are also characterized by some covariance matrix ${\bf{Q}}_{ij} = \mathbb{E}[p_i p_j]$.\footnote{The usual case of static ``long-only'' indicators is special since the corresponding correlation matrix is ill-defined. This will be the subject of a forthcoming work.} This matrix is in general non trivial, as one may systematically predict similar returns for two different assets $i$ and $j$, leading to ${\bf{Q}}_{ij} > 0$. In any case, one can as above build $N$ uncorrelated linear combinations of indicators, given by:
\be
\widehat p_i = \sum_j \left({\bf{Q}}^{-1/2}\right)_{ij} p_j,
\ee
with the above interpretation for ${\bf{Q}}^{-1/2}$. The $\hat p_i$'s are then all uncorrelated and of unit variance, i.e. with the same scale of predictability in all directions, 
and ``as close as possible'' to the original $p_i$'s, which is again financially meaningful. At this stage, any rotation in the space of (synthetic) assets also rotates the new indicators $\widehat p_i$ while keeping them all uncorrelated and of unit variance. The portfolio construction problem has thus become completely isotropic. 

\section*{Rotationally Invariant Portfolios}

How does all this help us to construct a truly agnostic Risk Parity portfolio, with no reference to a specific set of assets deemed fundamental? A simple observation is that the realized gain $\cal{G}$ of a portfolio invested in the synthetic asset $\alpha$ proportionally to $\widehat p_\alpha$ is given by:\footnote{This implicitly assumes that the cross-correlations between the $\widehat p_\alpha$ and the $\widehat r_{\beta \neq \alpha}$ are small, which is in fact an important hypothesis underlying our rotational symmetry principle.}
\be
{\cal G} = \sum_{\alpha=1}^N \widehat p_\alpha \cdot \widehat r_\alpha := \sum_{\alpha=1}^N {\cal G}_\alpha. 
\ee
This portfolio has several very desirable properties: 
\begin{itemize}
\item The risk associated with each synthetic asset is the same: $\mathbb{E}[{\cal G}_\alpha^2] = \mathbb{E}[\widehat p_\alpha^2]\mathbb{E}[\widehat r_\alpha^2]=1$, provided 
one neglects $\mathbb{E}[{\cal G}_\alpha]$ -- see footnote 1. 
\item The gains associated with different synthetic assets are uncorrelated: $\mathbb{E}[{\cal G}_\alpha \, {\cal G}_\beta]=\delta_{\alpha,\beta}$ -- see previous footnote 4. 
\item Most importantly, the total gain ${\cal G}$ is {\it invariant} under any further simultaneous rotation $\mathbb{R}$ of the assets and the indicators, as should be
for a scalar product:
\bea \nonumber
{\cal G}_R &=& \sum_{\alpha=1}^N \sum_{\beta=1}^N \mathbb{R}_{\alpha,\beta} \widehat p_\beta \cdot \sum_{\gamma=1}^N \mathbb{R}_{\alpha,\gamma} \widehat r_\gamma \\ \nonumber
&=&  \sum_{\beta=1}^N \sum_{\gamma=1}^N  \widehat p_\beta \cdot \widehat r_\gamma \sum_{\alpha=1}^N \mathbb{R}_{\alpha,\beta} \mathbb{R}_{\alpha,\gamma}\\
&=&  \sum_{\beta=1}^N  \widehat p_\beta \cdot \widehat r_\beta \equiv {\cal G}
\eea
where we have used the fundamental property of rotation matrices $\mathbb{R}\mathbb{R}^T = \mathbb{I}$. 
\end{itemize}
The last property means that {\it any arbitrary choice} of uncorrelated, unit variance synthetic assets with its corresponding set of indicators 
leads to the very same gain, so one does not need to decide on supposedly more fundamental investment factors. 

Why should one invest in the synthetic asset $\alpha$ proportionally to $\widehat p_\alpha$? On the basis of symmetry arguments, this is the only rational choice. All investments directions 
are made statistically equivalent, any other choice would correspond to an arbitrary breaking of isotropy. In the language of Markowitz optimisation, this corresponds to the optimal portfolio of synthetic assets when the expected future return of $\alpha$ is $S \; \widehat p_\alpha$, where the expected Sharpe ratio $S$ is independent of $\alpha$. Note that this in fact relies on the assumption that $\mathbb{E}[\widehat p_\alpha \, \widehat r_\beta]= S \delta_{\alpha,\beta}$, i.e. that at the level of uncorrelated factors, there is no significant cross-prediction left. This is, we believe, a very plausible assumption in practice -- see below.

Now, we need to convert the above isotropic risk portfolio invested in synthetic assets into tradeable contracts. This simply follows from the definition
of $\widehat r_\alpha$ and $\widehat p_\alpha$:
\bea \nonumber
{\cal G} &=& \sum_{\alpha=1}^N  \sum_{i,j=1}^N \left({\bf Q}^{-1/2}\right)_{\alpha j} p_j \left({\bf C}^{-1/2}\right)_{\alpha i} r_i \\ \nonumber
&=& \sum_{i=1}^N \left( \sum_{\alpha=1}^N \sum_{j=1}^N  \left({\bf C}^{-1/2}\right)_{\alpha i} \left({\bf Q}^{-1/2}\right)_{\alpha j} p_j \right) r_i \\
&:=& \sum_{i=1}^N \pi_i r_i,
\eea
where the last equation defines the physical position $\pi_i$ in a asset $i$, which is thus found to be:
\be
\pi_i =  \omega \sum_{\alpha=1}^N \sum_{j=1}^N  \left({\bf C}^{-1/2}\right)_{\alpha i} \left({\bf Q}^{-1/2}\right)_{\alpha j} p_j,
\ee
where $\omega$ is a constant that sets the overall risk of the portfolio, or, in vectorial form (using the symmetry of ${\bf C}$):

\be \label{result}
\boxed{
{{\bf \pi}} \;=\; \omega \, {\bf C}^{-1/2} \, {\bf Q}^{-1/2}  \;\; {\bf p}
}
\ee

This is the central result of this paper, that we now comment and specialize to several situations. First, let us notice that the above portfolio construction is such that the expected risk along any eigen-direction 
of ${\bf C}$ is the same, hence the name ``Eigenrisk Parity Portfolio'' (ERP) -- on this topic, see also \cite{PP,MaxEnt}. Indeed, the expected risk along the $a^{\text{th}}$ principal component is given by:
\be
{\cal R}_a = \mathbb{E}[ ({\bf \pi} \cdot {\bf v}_a )^2] \lambda_a,
\ee
where $\lambda_a$ is the $a^{\text{th}}$ eigenvalue of ${\bf C}$ and ${\bf v}_a$ the corresponding eigenvector. Simple algebra then leads to:
\be
{\cal R}_a = \omega^2 \left(\frac{1}{\sqrt{\lambda_a}}\right)^2 \mathbb{E}[ ({\bf v}_a \cdot {\bf Q}^{-1/2} {\bf p})^2 ] \lambda_a = \omega^2 \qquad \forall a,
\ee
where we have used the fact that the expected covariance of the indicator is ${\bf Q}$. Note that although for any given day the allocation 
${{\bf \pi}}$ points in a specific direction and is thus ``fully concentrated'' in that sense, this direction is expected to change over time -- provided the 
indicators themselves are not static. Isotropy is thus statistically restored on long enough time scales.

\section*{Agnostic Risk Parity}

Now, the naive choice for the indicator covariance matrix ${\bf Q}$ should be proportional to the return covariance matrix itself, i.e.  ${\bf Q} \propto {\bf C}$. In a stationary world where the indicators would really statistically predict future returns, i.e. $p_i=\mathbb{E}[r_i^\text{fut.}]$, this assumption would be natural, at least when ${\bf C}$ is computed on the time scale of the predicted returns, which is usually much longer than a day. Interestingly, plugging ${\bf Q} \propto {\bf C}$ in Eq. (\ref{result}) above precisely leads to the standard Markowitz optimal portfolio: ${\bf \pi} = \omega {\bf C}^{-1} \mathbb{E}[{\bf r}\, ^\text{fut.}]$.  However, this is a highly over-optimistic view of the world that only deals with ``known unknowns''. Directional predictions are extremely uncertain, much more so than risk predictions. In fact, directional predictions should not even be possible in an efficient market. If one insists that some signals may (weakly) predict future returns, it is wiser not to assume any particular structure on the correlation matrix of these indicators that any optimizer would use to hedge some bets with other bets. The most agnostic choice, less prone to unknown unknowns, is to choose ${\bf Q}=\sigma_{\text{p}} \mathbb{I}$, i.e. no reliable correlations between the realized predictions, and the same amount of predictability (or expected Sharpe ratio) on all assets. This leads to a very interesting portfolio construction:

\be
\boxed{
{\bf \pi}^* \;=\; \omega {\bf C}_{\text{RIE}}^{-1/2} \; {\bf p}
}
\ee

coined henceforth as ``Agnostic Risk Parity'' (ARP) because this specific asset allocation allows one to precisely balance the risk between all the principal components of the (cleaned) covariance matrix ${\bf C}_{\text{RIE}}$, in the worst-case scenario where the realized correlations between indicators would completely break down.   

\begin{figure} 
\begin{center}
\includegraphics[scale=0.7]{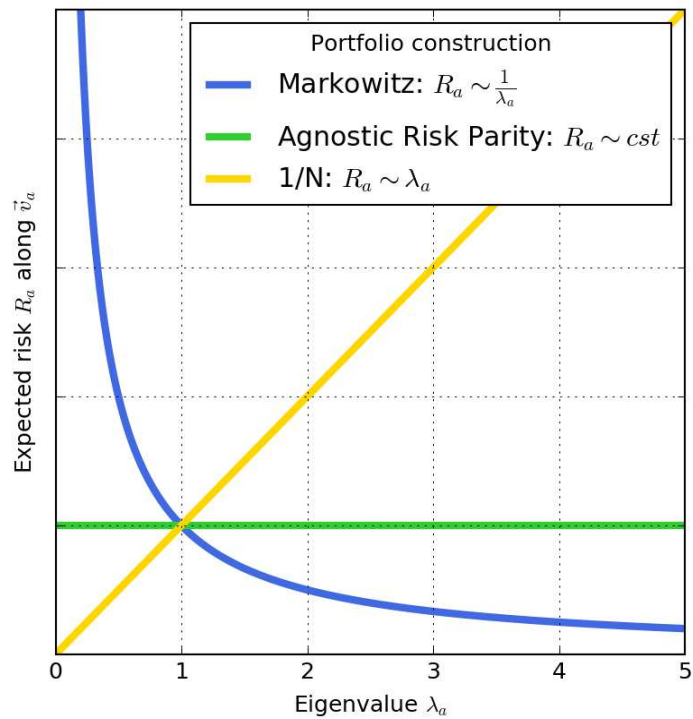}
\caption{Realized risk carried by different eigen-modes resulting from three portfolio constructions: 
$1/N$ on futures contracts, Markowitz, and Agnostic Risk Parity, all in the case where indicators are such that their realized covariance is ${\bf Q}=\sigma_{\text{p}}\mathbb{I}$.}
\label{fig_risk}
\end{center}
\end{figure}

Note that there is no explicit optimisation used in this argument -- rather, we look for a rotationally invariant portfolio construction with the minimal amount of information on the correlation structure of the indicators. The risk distribution per eigen-mode for various portfolio allocations is drawn in Fig. \ref{fig_risk}, when the realized covariance of the indicators is ${\bf Q} = \sigma_{\text{p}} \mathbb{I}$.  Note that, as is well known, the Markowitz optimisation scheme tends to over-allocate on small eigen-modes, which can lead to significant out-of-sample (bad) surprises \cite{Michaud}, a bias that is corrected within the ARP framework. 

Finally, one might believe that although uncertain, part of the return correlations could be inherited by the indicators. A simple way to encode this is to use for ${\bf Q}$ a shrinkage estimator, i.e. 
${\bf Q} \propto \varphi {\bf C}_{\text{RIE}} + (1 - \varphi) \mathbb{I}$, where $\varphi \in [0,1]$ allows one to smoothly interpolate between complete uncertainly ($\varphi=0$), corresponding to ARP, and the standard Markowitz prescription ($\varphi=1$). 

\section*{Agnostic Trend Following}

\begin{figure} 
\begin{center}
\includegraphics[scale=0.7]{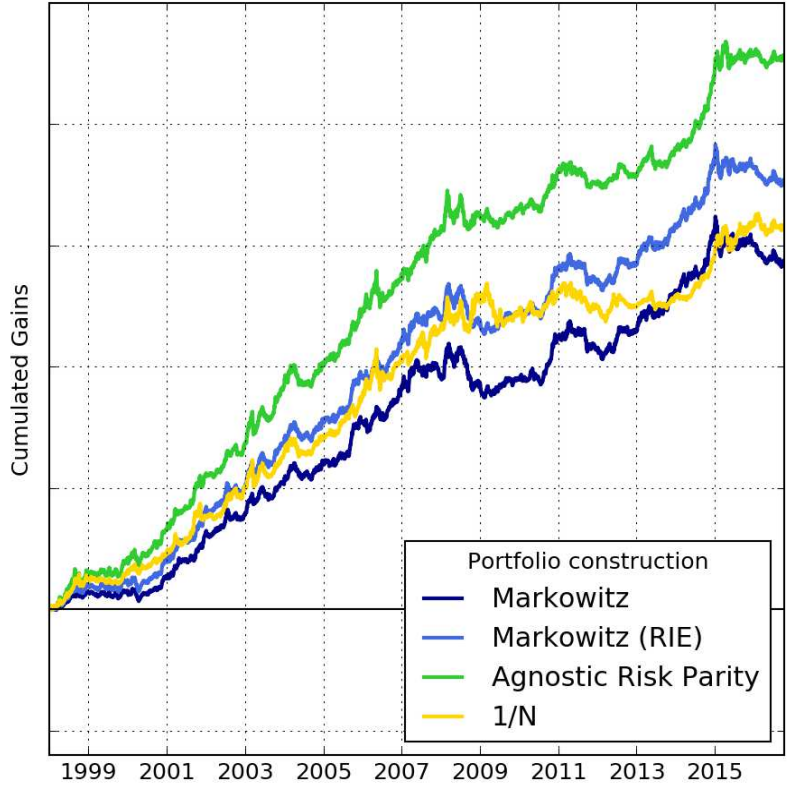}
\caption{Profit \& Loss (P\&L) curves for universal trend following for four portfolio constructions: $1/N$ on futures contracts, Markowitz with or without a cleaned RIE correlation matrix, and Agnostic Risk Parity, again with RIE. The universe here is composed of 110 contracts (commodities, FX, indices, bonds and interest rates). The trend indicator is a 1-year flat moving average of past returns. All P\&L's are rescaled such that their realized volatility is the same.}
\label{fig_pnls}
\end{center}
\end{figure}

The previous discussion was rather formal. As an example, we consider here the universal ``Trend'' indicator, based on a 1-year flat moving average of past returns of a collection of 110 futures contracts (commodities, FX, indices, bonds and interest rates) -- see the discussion in \cite{twohundred}. We normalize the returns of all futures and all the predictors to have unit variance. We then use three different portfolio constructions: equal $1/N$ risk on each physical asset, Markowitz optimal portfolio with either the raw empirical correlation matrix or a cleaned version ${\bf C}_{\text{RIE}}$ (using the RIE estimator detailed in \cite{RIE}, and no future information) and the Agnostic Risk Parity, again using the RIE estimator for ${\bf C}_{\text{RIE}}$. The P\&L's of the different portfolio since 1998 are shown in Figure \ref{fig_pnls}. While part of the improvement comes -- as expected -- from using a cleaned correlation matrix, we see that Agnostic Risk Parity yields the best result. Clearly the true correlation of predicted yearly returns ${\bf Q}$ is nearly impossible to measure without centuries of data, hence motivating the choice ${\bf Q} = \sigma_{\text{p}} \mathbb{I}$. We have observed similar results for other standard CTA strategies.

\section*{Perspectives}

In summary, we have offered a new perspective on portfolio allocation, which avoids any explicit optimisation but rather takes the point of view of {\it symmetry}. In a context where linear combinations of assets can easily 
be synthesized in a portfolio whose risk is measured through volatility, the asset space can be made fully ``isotropic'', in the sense that no preferred directions (corresponding to specific risk factors) can 
be identified. Therefore, in the absence of extra information, portfolio construction should respect this symmetry. This only requirement leads to a precise allocation formula, Eq. (\ref{result}), that generalizes Markowitz' prescription such as to take into account the expected correlation between the predicted returns of each asset in the portfolio. We have argued that the most agnostic choice, which is probably the most robust one out-of-sample, is 
to assume that these correlations are zero, i.e. that one should refrain from trying to hedge different bets if there is no certainty about the correlations between these bets. This leads to an Agnostic Parity Portfolio 
that realizes an equal risk over all principal components of the covariance matrix. We found that such an allocation over-performs Markowitz' portfolios when applied to classic technical (CTA) strategies, such as (universal) trend following.
There are several routes that should be explored further. For example, non-quadratic measures of risk, such as skewness or kurtosis, would break rotational symmetry and possibly lead to meaningful fundamental risk factors that should be maximally diversified (see e.g. \cite{Higher}). We leave this for future work.  

\vskip 1cm

We thank N. Bercot, J. Bun, R. Chicheportiche, S. Ciliberti, C. Deremble, L. Duchayne, L. Laloux, A. Rej for many useful discussions on these issues. 

\section*{Appendix}

We consider random variables ${\bf r}$ of mean zero and unit variance, with a correlation matrix given by ${\bf C}$. We are looking for $\widehat {\bf r}$, a linear transformation of ${\bf r}$ such that $\mathbb{E}[\widehat r_i \widehat r_j] = \delta_{ij}$. Clearly $\widehat {\bf r} = {{\bf C}}^{-1/2} {\bf r}$ satisfy this property, and any solution is of the form $\widehat {\bf r} = \mathbb{R} {{\bf C}}^{-1/2} {\bf r}$ where $\mathbb{R}$ is a rotation matrix. We further demand that the following Mahalanobis distance $d(\mathbb{R})$ is minimized:
\be
d(\mathbb{R}) = \mathbb{E}\left[\left(\widehat {\bf r} - {\bf r}\right) \cdot {{\bf C}}^{-1} \left(\widehat {\bf r} - {\bf r}\right)\right].
\ee
Expanding the square, one readily sees that the only term that depends on $\mathbb{R}$ is $-2 \text{Tr}[\mathbb{R}{{\bf C}}^{-1/2}]$, which must be maximized. Since ${{\bf C}}^{-1/2}$ is a positive
definite matrix, it is immediate to show that the optimal solution is $\mathbb{R}=\mathbb{I}$, i.e. ${{\bf C}}^{-1/2}$ is diagonal in the basis of ${{\bf C}}$. Although natural in the present context, we note that changing the Mahalanobis distance to any other positive definite quadratic form where ${{\bf C}}^{-1}$ is replaced by any (matrix) function of ${{\bf C}}$ -- including the identity matrix -- leads to the same result.

\end{document}